\documentclass[a4paper,11pt]{article}
\usepackage[T1]{fontenc}
\usepackage[utf8]{inputenc}
\usepackage{graphicx}
\usepackage{multicol}

\usepackage{hyperref}
\hypersetup{
  setpagesize  = false,
  colorlinks   = true,    
  urlcolor     = blue,    
  linkcolor    = black,   
  citecolor    = black    
}
\usepackage[outercaption]{sidecap}

\usepackage{authblk}
\usepackage{geometry}
\usepackage{setspace}
\usepackage{verbatim}
\usepackage[utf8]{inputenc}
\usepackage{amsmath}
\usepackage{amssymb}
\usepackage{gensymb}
\usepackage{verbatim} 
\usepackage{amsfonts}
\usepackage{cancel}

\usepackage{marginnote}

\usepackage[
    backend=biber,
    style=ieee,
  ]{biblatex}

\newcommand{\keywordsenglishname}{Keywords}

\renewenvironment{abstract}{%
        \begin{center}
	\begin{minipage}{14cm}
	{\textbf{\abstractname:}}
}{
        \end{minipage}
	\end{center}
}
\newenvironment{abstractinenglish}{
        \def\abstractname{\abstractinenglishname}
	\begin{abstract}
}{
        \end{abstract}
}

\newenvironment{keywordsenglish}{
        \def\abstractname{\emph{\keywordsenglishname}}
	\begin{abstract}
}{
        \end{abstract}
}

\title {Blue laser induced bright red fluorescence in hot cesium vapor }
\author{Armen Sargsyan, Anahit Gogyan$^{\dagger}$, and David Sarkisyan}
\affil{Institute for Physical Research, National Academy of Sciences of Armenia, Gitavan-2, 0204 Ashtarak, Armenia}
\affil{\small{$^\dagger$ Corr. author: agogyan@gmail.com}}
\date{}
\usepackage{fancyhdr}
\begin{document}

\maketitle
\vspace{6pt}

\begin{abstractinenglish}
We have observed laser-induced fluorescence using 456~nm laser radiation, resonant with the 6S$_{1/2}$–7P$_{3/2}$ transition in Cs atoms. It includes red emission lines in the range of 580-730~nm and a prominent line at 852~nm corresponding to the 6P$_{3/2}$-6S$_{1/2}$ transition. A T-shaped all-sapphire cell with a length of 1~cm, containing Cs atomic vapor and capable of being heated up to 500$^\circ$C, was used. 
The laser-induced fluorescence (LIF) power at 852~nm was investigated as a function of the cell temperature. The maximum LIF power was achieved at 130$^\circ$C, while a significant decrease was observed around 300$^\circ$C. At 130$^\circ$C, the Doppler-broadened LIF spectrum at 852~nm exhibited self-conversion, resulting in the formation of two distinct peaks within the spectrum. The LIF power at 852~nm was also studied as a function of the 456~nm radiation power.
The Cs cell demonstrated potential as an efficient optical filter and down-converter, effectively transforming 456~nm radiation into 852~nm radiation.

\end{abstractinenglish}

\begin{keywordsenglish}
 laser-induced fluorescence;  hot cesium vapor;  down-converter   
\end{keywordsenglish}
 
\vspace{6pt}
\section{Introduction}\label{intro}
In \cite{ref1}, it was experimentally demonstrated that blue laser radiation could be successfully used for wireless underwater communication in free space over a range of 36 kilometers in the Yellow Sea. Particularly, blue laser radiation at 456~nm exhibits a relatively small absorption coefficient of 0.01~m$^{-1}$ in certain ocean waters \cite{ref2}. Therefore, it is crucial to study the conditions for effective detection and manipulation of blue radiation conversion. By registering 852~nm laser-induced fluorescence (LIF), it is possible to completely suppress 456~nm radiation, enabling a cesium (Cs) vapor cell to serve as an optical filter \cite{ref3, ref4}.

Currently, the Cs atom transitions 6S$_{1/2} \to$ 7P$_{1/2, 3/2}$ (D$_{1,2}$ lines) at wavelengths 459~nm and 456~nm, respectively, are actively being studied. Nonlinear magneto-optical resonances observed in the fluorescence to the ground state from the  7P$_{3/2}$ level, populated directly by 456~nm laser radiation, were investigated in \cite{ref5}. The Cs D$_2$ transition was theoretically analyzed in \cite{ref6} and compared with the D$_1$ transition (\(\lambda = 459~\mathrm{nm}\)). In \cite{ref7}, the D$_2$ line transition was studied using the saturated absorption (SA) technique, allowing for sub-Doppler spectroscopy. The same technique was applied to examine the D$_1$ transition \cite{ref8}.

Population inversion between the 7S$_{1/2}$ and 6P$_{3/2}$ levels of Cs in a thermal cesium cell, achieved using a 455.5~nm pumping laser, and the subsequent laser generation at 1470~nm were demonstrated in \cite{ref9}. Moreover, by using 456~nm radiation in combination with additional 1070~nm radiation, high-lying Rydberg levels such as 32S$_{1/2}$ can be explored, as shown in \cite{ref4}. A bichromatic pumping scheme with 852- and 917-nm lasers excites Cs atoms to the 6D$_{5/2}$ level, followed by cascaded decay, resulting in coherent 456~nm blue laser generation under phase-matching conditions \cite{ref10, ref11}. Generation of polychromatic and collimated light at 456~nm, 459~nm, and 761~nm via two-photon excitation of the 6S$_{1/2}$ $\rightarrow$ 8S$_{1/2}$ transition with two infrared pump lasers at 852~nm and 795~nm is demonstrated in \cite{ref12, ref13}. It is noteworthy that \cite{ref10} mentions that 456~nm generation via two-photon excitation of the 6S$_{1/2}$ $\rightarrow$ 6D$_{5/2}$ transition may be more advantageous than using two-photon excitation of the 6S$_{1/2}$ $\rightarrow$ 8S$_{1/2}$ transition.

In this work, laser radiation at 456~nm, which is in resonance with the 6S$_{1/2}$ $\rightarrow$ 7P$_{3/2}$ transition, generates bright red laser-induced fluorescence  and a strong 852~nm line corresponding to the 6S$_{1/2}$ $\rightarrow$ 6P$_{3/2}$ transition.
In contrast to some of the aforementioned works, this study demonstrates a straightforward scheme for converting blue radiation into red using only a single 456~nm laser, without requiring the use of two lasers.

\begin{figure}[tb] 
    \centering
    \includegraphics[width=0.45\textwidth]{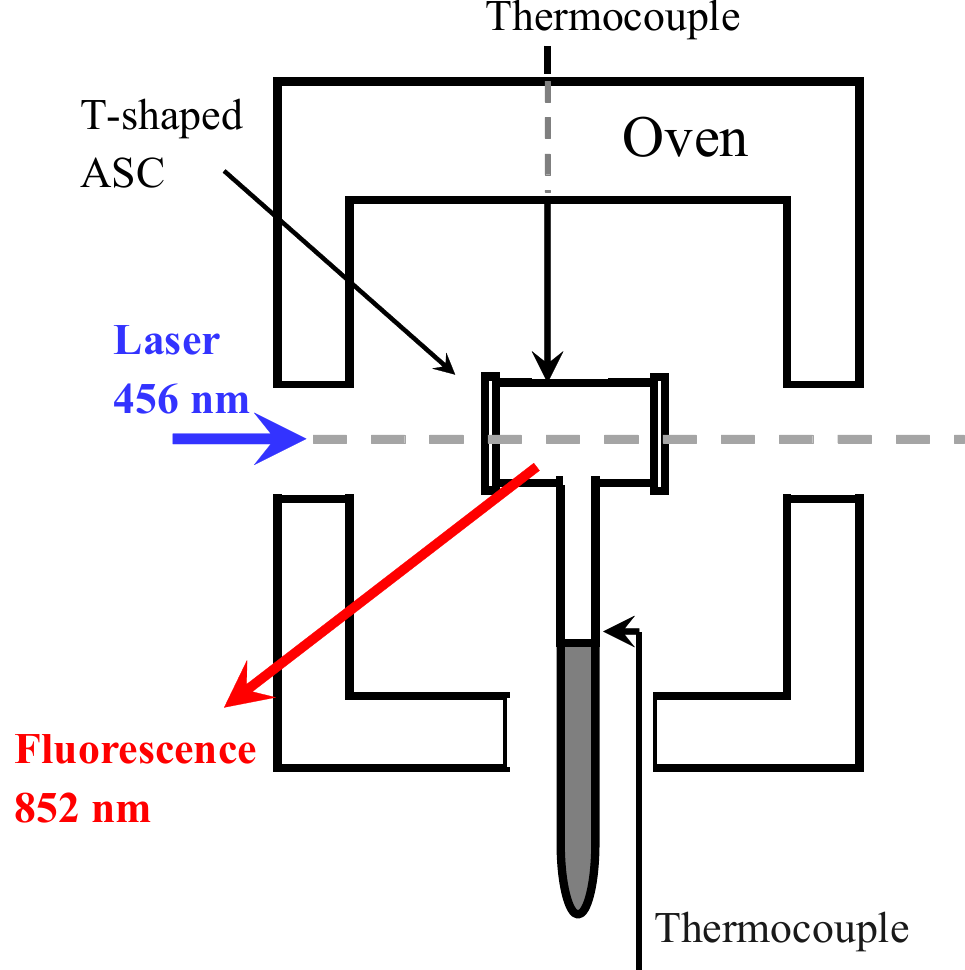}
    \caption{
        A homemade T-shaped 1-cm long all-sapphire cell containing Cs atomic vapor, housed within a heater. An ECDL laser with a spectral linewidth of approximately 0.4~MHz, a beam diameter of 2~mm, and tunable around $\lambda = 456$~nm was used for resonant excitation of the $6S_{1/2} \rightarrow 7P_{3/2}$ transitions in Cs atoms. Fluorescence signal at 852~nm is registered with a photodetector. 
    }
    \label{fig:setup}
\end{figure}

An all-sapphire cell (ASC) of 1-cm length containing Cs atomic vapor and capable of being heated up to 500$^\circ$C is utilized \cite{ref14}. The LIF power of 852~nm radiation is studied as a function of the cell temperature, reaching a maximum at 130$^\circ$C and strongly decreasing around 300$^\circ$C. Meanwhile, LIF at 890~nm corresponding to the 6S$_{1/2}$ $\rightarrow$ 6P$_{1/2}$ transition is detected with three times less power. The LIF intensity increases nearly linearly with 456~nm laser power up to 100~mW, achieving a conversion efficiency of 1\%. Using 456~nm radiation in resonance with the 6S$_{1/2}$ $\rightarrow$ 7P$_{3/2}$ transition, the dependence of bright red LIF and 852~nm LIF on ASC temperature and laser intensity is thoroughly investigated.

The next section presents the experimental details and analyzes the obtained results, while the Conclusion summarizes the findings.

\begin{figure}[tb] 
    \centering
    \includegraphics[width=0.45\textwidth]{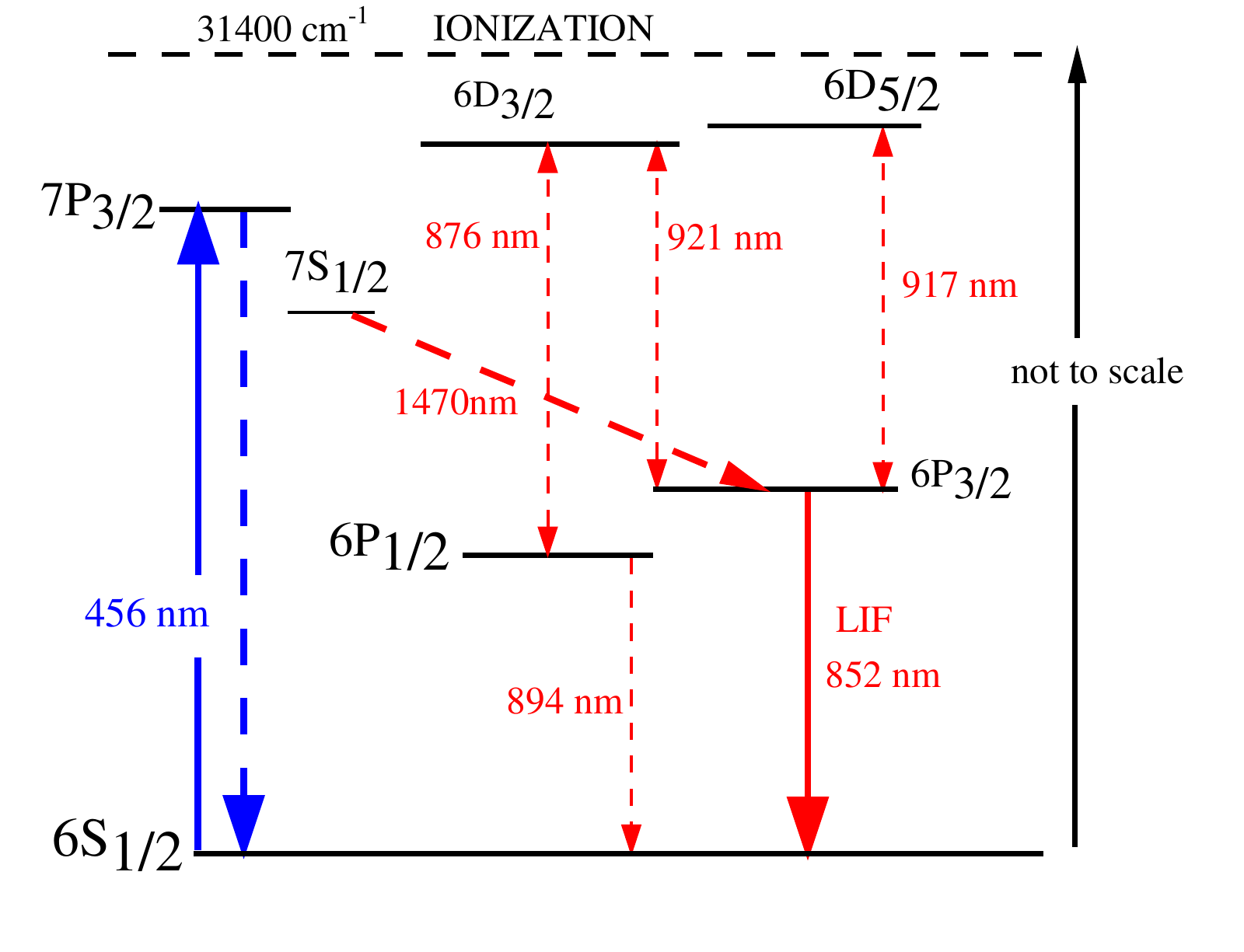} 
    \caption{
        Energy level diagram of atomic cesium, highlighting the states relevant to this study. The excitation of the $6S_{1/2} \rightarrow 7P_{3/2}$ transition occurs at 456~nm. LIF is observed at 456~nm, 894~nm, and 852~nm.
    }
    \label{fig:energy_levels}
\end{figure}

\begin{figure}[tb] 
    \centering
    \includegraphics[height=0.3\textwidth]{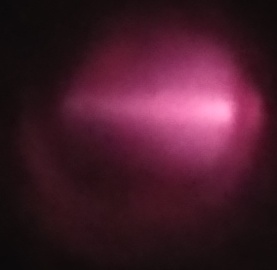} 
    \hspace{0.05\textwidth}
    \includegraphics[height=0.3\textwidth]{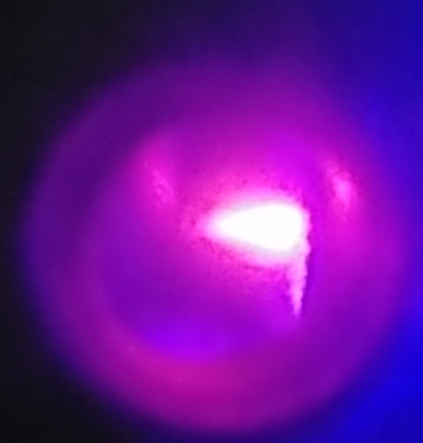} 
    \caption{
        Excitation of the $6S_{1/2} \rightarrow 7P_{3/2}$ transition using 100 mW of 456~nm radiation. A strong LIF of bright red color, consisting of several prominent lines in the range of 580–730~nm and at 852~nm, is detected using a conventional photo camera. The left image corresponds to the ASC temperature of $\sim 100^\circ$C, while the right image corresponds to $\sim 130^\circ$C. The bright white spot at the center of the red region is due to the saturation of the photo camera's sensitivity.
    }
    \label{fig:LIF_red}
\end{figure}

\begin{figure}[tb] 
    \centering
    \includegraphics[width=0.48\textwidth]{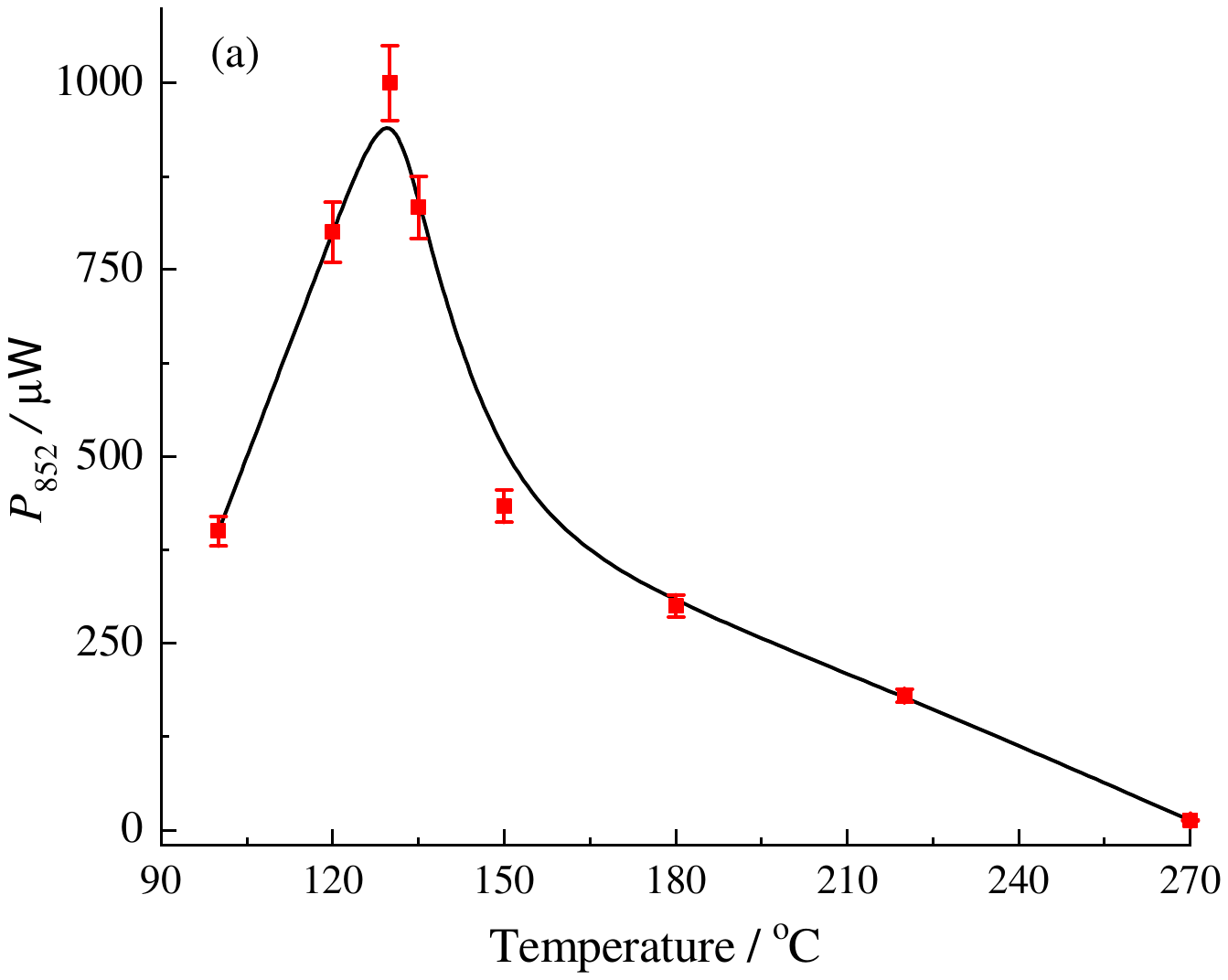}
    \includegraphics[width=0.48\textwidth]{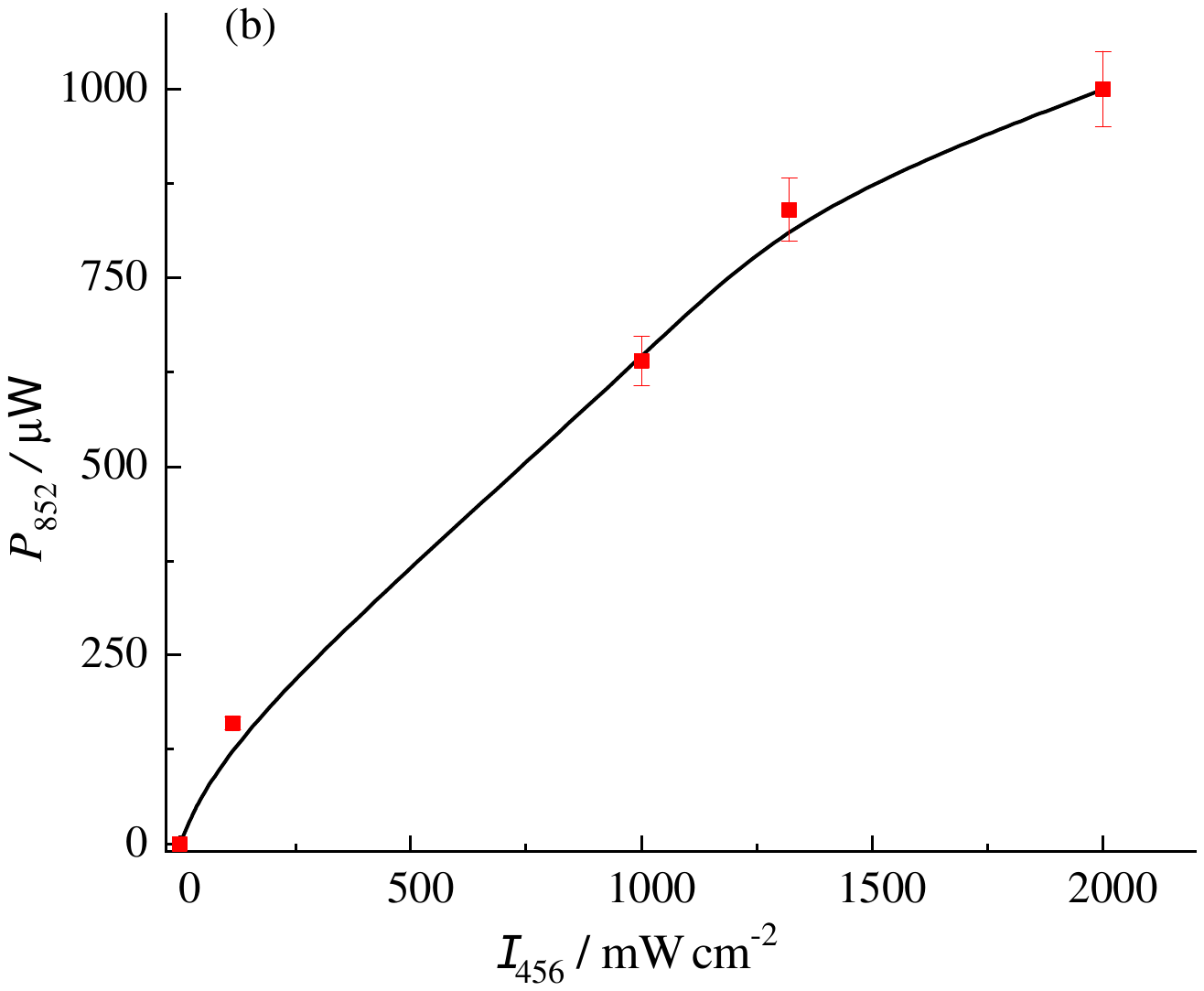}
        \caption{(a) Dependence of LIF power at 852~nm as a function of ASC temperature. (b) LIF power at 852~nm as a function of incident radiation intensity at 456~nm, measured at an ASC temperature of 130~$^\circ$C.
   }
    \label{fig:power_dependence}
\end{figure}

\begin{figure}[htb] 
    \centering
    \includegraphics[width=0.55\textwidth]{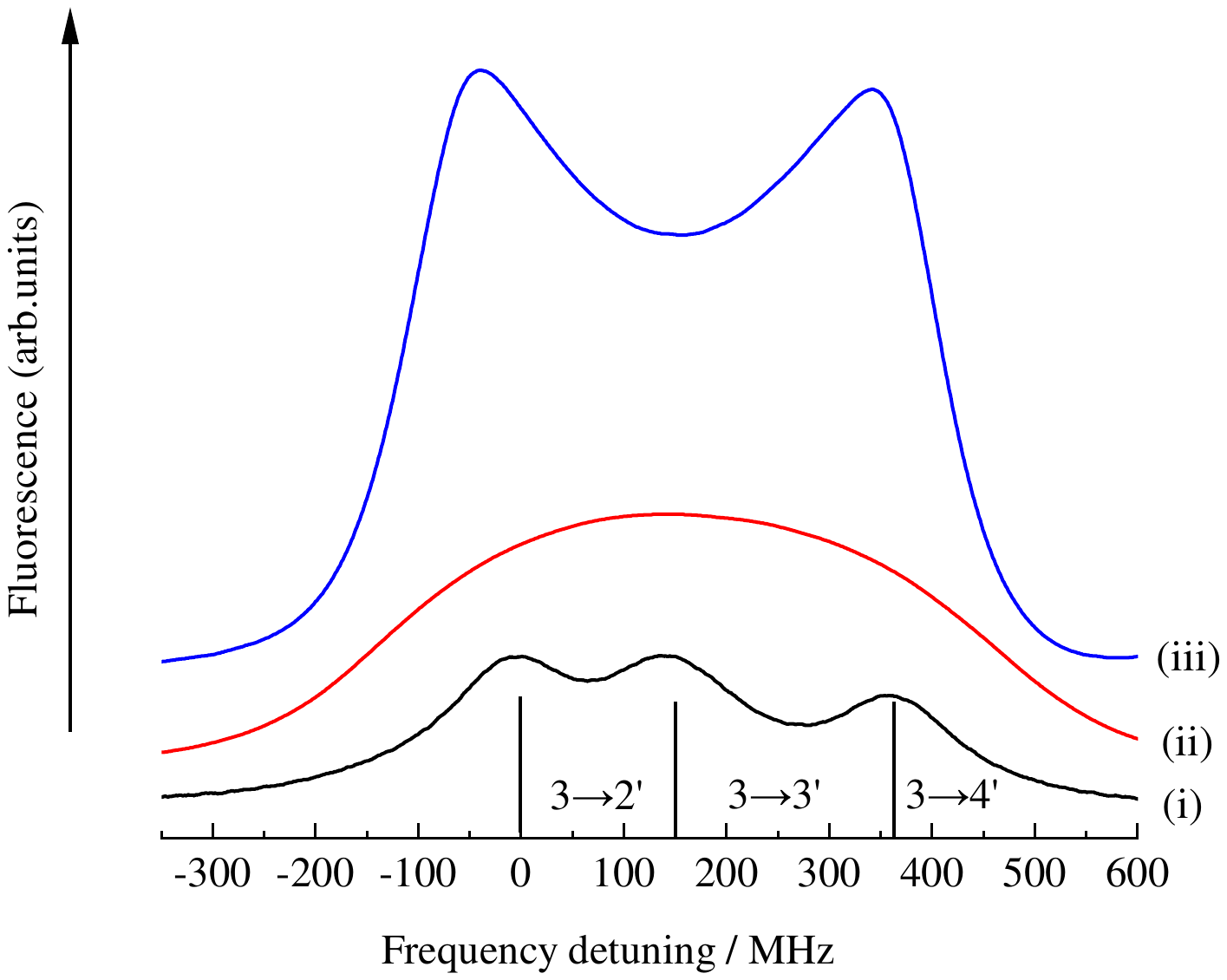} 
        \caption{
        Fluorescence spectra of cesium under different conditions. 
        Curve (i): Sub-Doppler reference fluorescence spectrum of the Cs D$_2$ line ($F_g=3 \rightarrow F_e=2,3,4$ transitions) obtained using a Cs nanocell heated to $100^\circ$C with an atomic vapor column thickness of $L=850$~nm. 
        Curve (ii): Doppler-broadened fluorescence spectrum at 852~nm recorded at $100^\circ$C. 
        Curve (iii): Doppler-broadened fluorescence spectrum at 852~nm recorded at $130^\circ$C, demonstrating self-conversion and forming two peaks inside the spectrum.
    }
    \label{fig:Cs_fluorescence_spectra}
\end{figure}

\section{Experimental Procedures and Results}\label{exp}

\subsection*{Experiment}

A part of the experimental setup is sketched in Fig.~1. A tunable external-cavity diode laser with a spectral linewidth of $\sim$0.4~MHz, beam diameter of 2~mm, and tunable around $\lambda = 456$~nm was used for the resonant excitation of the $6S_{1/2} \rightarrow 7P_{3/2}$ transitions of Cs atoms. The laser beam was directed normally to the windows of a homemade T-shaped 1-cm-long ASC containing Cs atomic vapor, as shown in Fig.~1. 

The specially designed oven consists of two heaters: one for the cell body and another for the side-arm (reservoir) containing Cs metal vapors. Due to its original construction, the pressure of the Cs atomic vapor is determined by the temperature at the boundary of the Cs metal column in the side-arm. The oven has three openings: two for laser transmission and one for LIF detection in a direction perpendicular to the laser propagation.

The atomic Cs levels and the wavelengths of transitions produced by the 456~nm laser are shown in Fig.~2. The 456~nm radiation effectively populates the $7P_{3/2}$ level (energy $\sim$21,950~cm$^{-1}$), from which the Cs atom absorbs another 456~nm photon, leading to ionization. Subsequent recombination results in the Cs atom occupying higher energy levels, which, through cascade spontaneous emission, predominantly populate the $6P_{3/2}$ level~\cite{ref3, ref4}. Narrowband 
interference filters were used to select LIF at 456~nm, 852~nm, or 894~nm. The power of these radiations was measured 
by a power meter and recorded with a photodiode FD-24K and a four-channel Tektronix TDS2014B 
oscilloscope.

\subsection*{Results}
A strong LIF of bright red color, consisting of several intense red lines in the range of 580--730~nm and the 852~nm line, was observed at ASC temperatures of $\sim$100~$^{\circ}$C and $\sim$130~$^{\circ}$C (Fig.~3, left and right panels). These were detected using a conventional photo camera. The Cs vapor densities were $N = 1.6 \times 10^{13}$~cm$^{-3}$ and $N = 8 \times 10^{13}$~cm$^{-3}$, respectively. Notably, when the 456~nm radiation was detuned by 300~MHz from the $6S_{1/2} \rightarrow 7P_{3/2}$ transition, the red LIF disappeared. This makes it convenient to visually ensure resonance of the laser frequency with the $6S_{1/2} \rightarrow 7P_{3/2}$ transition by maximizing the brightness of the red radiation, which is clearly visible to the naked eye even from several meters away.Notably, blue fluorescence at 456~nm is also observed in the emitted fluorescence light.

The dependence of LIF power at 852~nm on ASC temperature is shown in Fig.~4(a). The maximum power was achieved at $130^{\circ}$C, while at higher temperatures it decreased. Recalculating the total 852~nm power emitted into $4\pi$ steradians (i.e., spontaneously emitted in all directions) indicates a conversion efficiency of the 456~nm radiation power of $\sim$1\%. The dependence of the LIF power at 852~nm on 456~nm radiation intensity at $130^{\circ}$C is shown in Fig.~4(b).

The spectrum of 852~nm radiation in the side direction, recorded using a photodiode and an oscilloscope, is shown in Fig.~5. Note that the sensitivity of the photodiode at 852~nm is seven times greater than at 456~nm, an additional benefit of the 456~nm~$\rightarrow$~852~nm conversion. Curve~(i) shows a sub-Doppler reference fluorescence spectrum of the Cs D$_2$ line $F_g=3 \rightarrow F_e=2,3,4$ transitions obtained with a Cs nanocell heated to $100^{\circ}$C, with an atomic vapor column thickness of $L = 850$~nm~\cite{ref15, ref16}. Curve~(ii) shows the 852~nm Doppler-broadened fluorescence spectrum at $100^{\circ}$C with a narrowband interference filter at 852~nm, and Curve~(iii) shows the spectrum at $130^{\circ}$C.

The broadened fluorescence spectrum at 852~nm demonstrates self-conversion, forming two peaks within the spectrum. The fluorescence dip at the spectrum center is due to absorption while passing through the atomic vapor column ($L = 0.5$~cm) before exiting the cell. The absorption is approximately $\exp(-\sigma NL)$~\cite{ref17}, where $\sigma = 10^{-11}$~cm$^2$ is the Cs absorption cross-section at the exact $6S \rightarrow 6P_{3/2}$ transition, $N$ is the Cs density, and $L$ is the vapor column length. Here, $\sigma NL \sim 400$, although this {value might be smaller by an order of magnitude, as many Cs atoms are in excited states when using 456~nm excitation}~\cite{ref9}. Note that $\sim$8\% LIF efficiency is observed for 456~nm radiation.

\section{Conclusion}

Using 456~nm laser radiation resonant with the $6S_{1/2} \rightarrow 7P_{3/2}$ transition of Cs atoms and a T-shaped ASC containing Cs atomic vapor (capable of being heated up to 500~$^{\circ}$C), a bright LIF was observed at red lines (in the range of 580--730~nm) and at the 852~nm line ($6P_{3/2} \rightarrow 6S_{1/2}$ transition). The dependence of LIF power at 852~nm on ASC temperature was studied, with a maximum power achieved at $130~^{\circ}$C, which significantly decreased at $\sim$300~$^{\circ}$C. LIF at the 894~nm transition (D$_1$ line) was also detected, but with approximately three times less power.

At $130~^{\circ}$C, the Doppler-broadened LIF at 852~nm demonstrated self-conversion, forming two peaks in the spectrum. This behavior was attributed to Cs vapor absorption as the radiation propagated toward the exit of the ASC. The 852~nm LIF power exhibited a nearly linear dependence on the 456~nm laser intensity, up to 100~mW. The ASC was repeatedly heated to $300~^{\circ}$C and higher without any darkening caused by the interaction of hot Cs vapors, in contrast to glass cells, where darkening occurs at temperatures above $150~^{\circ}$C~\cite{ref14}. Recalculating the total 852~nm power, which is spontaneously emitted in all directions, revealed a down-conversion efficiency of 456~nm~$\rightarrow$~852~nm of $\sim$1\%. Furthermore, at the input to the photodetector, the 456~nm radiation could be completely suppressed, indicating that the ASC could serve as an efficient 456~nm~$\rightarrow$~852~nm optical filter and down-converter.

To achieve exact resonance of the 456~nm laser with the $6S_{1/2} \rightarrow 7P_{3/2}$ transition, a simple and practical method is to visually maximize the brightness of the red radiation, which is clearly visible to the naked eye even from several meters away. This process, illustrated in Fig.~3, provides an excellent demonstration of the frequency down-conversion effect for students. For simplicity, a Cs glass cell heated to $130~^{\circ}$C could be used for this demonstration.

\section*{Credit Authorship Contribution Statement}

\textbf{Armen Sargsyan:} Investigation, Writing -- review and editing, Funding acquisition. \\
\textbf{Anahit Gogyan:} Formal analysis, Software. \\
\textbf{David Sarkisyan:} Writing -- original draft, Writing -- review and editing, Supervision, Funding acquisition. \\

\section*{Funding}

This work was supported by the Higher Education and Science Committee of the Republic of Armenia under the research project N~22IRF-06  and project N~1-6/23-I/IPR.

\printbibliography

\end{document}